# Analysis of SQL Injection Detection Techniques

Jai Puneet Singh *
CIISE, Concordia University, Montréal, Québec, Canada
`ja_ngh@live.concordia.ca`

*Abstract*—SQL Injection is one of the vulnerabilities in OWASP's Top Ten List for Web Based Application Exploitation. These types of attacks takes place on Dynamic Web applications as they interact with the databases for the various operations. Current Content Management System like Drupal, Joomla or Wordpress have all the information stored in their databases. A single intrusion into these types of websites can lead to overall control of websites by the attacker. Researchers are aware of the basic SQL Injection attacks but there are numerous SQL Injection attacks which are yet to be Prevented and Detected. Over here, we present the extensive review for the Advanced SQL Injection attack such as Fast Flux Sql Injection, Compounded SQL Injection and Deep Blind SQL Injection. We also analyze the detection and prevention using the classical methods as well as modern approaches. We will be discussing the Comparative Evaluation for prevention of SQL Injection.

## I. INTRODUCTION

Web Application are widespread today as they have become the necessity for the everyday life. There are thousands of security breaches that take place every day. According to Bagchi [1] 75% of the firms websites and web applications were vulnerable to the Internet Security Breaches. He had analyzed through Gompertz model the growth of Internet Security breaches and the vulnerability of an attack. The most common attack on web is through SQL Injection. The classical SQL Injections were easy to prevent and detect and a lot of procedures, methodologies were discussed in order to overcome SQL Injections. The various methodologies used to overcome the attack is by writing secure code according to the extensive research by howard and his team which relates to the writing of defensive code with proper validation by the usage of encoding and decoding techniques[2]. Still today, writing defensive codes is encouraged but it is not enough to protect SQL Injection Attacks. The SQLIA's are widespread attacks on the websites which are followed by the XSS (Cross Site Scripting) attack. A study by Gartner Group over 300 Internet Web sites has shown that most of them could be vulnerable to SQLIAs [3].

There are numerous types of SQLIA's and each has different approach for attacks onto the website. The complex formation occurs with the combination of SQL Injection and XSS attacks which lead to retrieval of the Database information. Even the SQL Injection attacks are taking place in the Rich Internet Application by finding the vulnerability in cross domain policies. Most of the modern websites extensively use Rich Internet Application[4] such as Adobe Flash and Microsoft Silver light, for increased user defined functionality. If the care is not taken during the coding of cross site scripts,it can lead to the vulnerability of XSS and SQL Injection Attacks. These types of attacks were not present a few years ago with the advancement in the field of UI/UX design and various other technological changes such as the introduction to JSON, JQuery which resulted in new vulnerabilities. So, In order to counter these attacks we will be extensively discussing the modern SQL Injection attacks and the ways to protect and defend these type of attacks.The negligence at the initial stage can lead to monetary losses at later stage.

The rest of the paper is organized as follows: Section 2 describes the Background of the SQL Injection Attacks and the concepts related to it. Section 3 details the example application which will be used throughout the paper for the discussion of the advanced SQL Injection Attacks (SQLIA). Section 4 lists the Advanced SQL Injection Attacks. Section 5 presents the techniques for the prevention and detection of such attacks. Section 6 gives out the Technique Evaluation in order to compare various detection and prevention mechanism. We provide the summary and conclusion in Section 7.

## II. BACKGROUND

SQL Injection has been studied for a period *SQL Injection Attack* and it occurs when the attacker tries to insert malicious code into the Web Application database which is intended for the retrieval or corruption of data. These attacks are moreover used on E-Commerce websites for the extraction of credit card numbers or it is widely used for bypassing the authentication. Su and Wassermann describe SQL Injection thoroughly and formally with an explanation on Code Injection as well as validation using SQL Check [5]. For website fields without proper input validation, an attacker could obtain direct access to the database of the underlying application [6].

## III. SQL INJECTION ATTACK TYPES

There are numerous research papers which present various SQL Injection attacks but most of them discuss the classical SQL Injection whereas the modern SQL Injection attack are more dangerous. The modern SQL Injection attack can overcome many previously discussed Detection and Prevention technique.This section is divided into two subsection which is allocated into Classical SQL Injection and Advanced SQL Injection.

### A. Classical SQL Injection

In this section we just give small overview of the classical sql injection which are as follows:

*1) Piggy Backed Queries:*
*Intent of Attack: Retrieval of Information, Denial of Service*
In this type of attack the attacker "Piggy Back" the query with the original query in the input fields present on the web application.The piggy backed can be defined as "on or as if on the back of another". The database receives the multiple queries [3]. During the execution, the premier query works as in a normal case the second query adjoined with the first query is used for SQL Injection Attack.It is considered as menacing attack since it fully exploits the database. Using proper prevention and detection technique this type of attack can be prevented. As an Example:

```
select customer_info from accounts where
login_id = ''admin'' AND pass = '123' ;
DELETE FROM accounts WHERE CustomerName =
'Albert';
```

After executing the first query the interpreter sees the ';' Semi Colon and executes the second query with the first query.The second query is malicious so it will delete the all the data of the customer 'Albert'. Hence, these type of malicious act can be protected by firstly determining the correct SQL Query through proper validation or to use different detection techniques. This type of attack can be prevented using Static Analysis, Run time monitoring is not needed.

*2) Stored Procedure:*
*Intent of Attack: Escaping Authentication, Denial of Service, More freedom on Database* Stored Procedures are widely used as a subroutine in a relational database management system. They are compiled into single execution plan and extensively used for performing commonly occurring tasks.Its used in businesses as it provides single point of control while performing the business rules.IT Professionals think that SQL Stored Procedures are remedy for the SQL Injection as Stored Procedures are placed on the front of the databases the security features cannot be applied to them.The stored procedures do not use the standard Structured Query Language, it uses its own scripting languages which does not have same vulnerability as SQL but different vulnerability related to the Scripting language still exist.As an Example:

```
CREATE PROCEDURE User_info
@username varchar2 @pass varchar2
@customerid int
AS
BEGIN
EXEC('Select customer_info from
customer_table where username='
''+@username '' ' and pass = ' ''+@pass
'' '
GO
```

Fig. 1: Stored Procedure Description

This type of procedures are vulnerable for the SQL Injection Attack. Any malicious user can enter the malicious data in the fields of username and password. The simple command entered by the user can destroy whole database or it can lead to service disruption. It is always advised, not to store the critical information on the stored procedures, as it lacks the most important security features.

*3) Union Query:* This type of attack uses Union Operator (U) while inserting the SQL Query. The two sql query are joined with the Union Operator. The first statement is a normal query after which the malicious query is appended with union operator. Hence, it is used to bypass the prevention and detection mechanism of the system.The example shows how it can be proceeded. The example shows that the second query is malicious and text following (–) is disregarded as it becomes comment for the SQL Parser. Taking this an advantage the attacker attacks the web application or website with this query.

```
select * from accounts where id='212'
UNION select * from credit_card where
user='admin'--' and pass='pass'
```

Fig. 2: The SQLI Attack with the Union Query

*4) Alternative Encoding:* In this type the attacker changes the pattern of the SQL Injection so that it goes undetected from the common detection and prevention techniques.In this method the attacker uses hexadecimal, Unicode , octal and ASCII code representation in the SQL Statement. It goes undetected with the common detection and prevention.As these could not be able to detect the encoded strings and hence, allows these attacks to go undetected.

*B. Advanced SQL Injection*

*1) Blind SQL Injection Attack: Intent of Attack:* Many Web applications disable to display the mysql or any other sql error messages.In this attack the information is inferred by asking true/false questions.If the injection point is completely blind then only way to attack is by using the WAIT FOR DELAY or BENCHMARK command. This type of injection is known as Deep Blind SQL Injection Attack [7].

*2) Fast Flux SQL Injection Attack:*
*Intent of Attack:Data Extraction, phishing*
Phishing is a significant security threat to the users of an Internet.The phishing is a social engineering attack in which an attacker fraudulently acquire sensitive information from the user by impersonating as a third party [8].Traditional phishing host can be detected very easily just by tracking down the public Domain Name Server or the IP address.This trace back technique could lead to the shutdown of the hosting websites.The attackers understood that conducting an attack this large could have significant effect on load balancing of a server [9].To counter this action in order to protect its criminal assets, the operator of phishing websites started using Fast Flux technique.Fast Flux is a Domain Name Server technique to hide phishing and malware distribution sites behind an ever changing network of compromised host. The fast flux attacking technique can be understood by the below diagram.

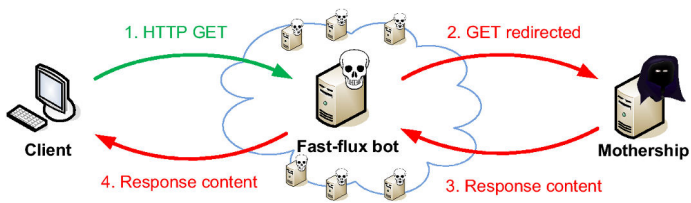

Fig. 3: Fast Flux Attack

The massive SQL Injection attack using fast flux came using the Asprox botnet.In Fast Flux mode, the DNS(Domain Name server) simultaneously hosts many different malware infected IP's and the IP's constantly changing. The first fast flux SQL Injection was detected in banner82.com which now has been closed but it was studied by researchers thoroughly.It was infecting new hosts to be added to the botnet. Banner82.com (now a closed domain) has a tiny iFrame that's attempting to load dll64 .com /cgi-bin/index.cgi?admin where the NeoSploit malware exploitation kit is serving MDAC ActiveX code execution (CVE-2006-0003) expl [10]. The figure 2. and figure 3. shows that in Fast Flux Attacks the IP address is continuously changing so it is very difficult to trace down the website and to stop from spreading the malware.The fast flux attacks in SQL Injection is difficult to predict and defend. Therefore, a very little research for protection and detection has been done for the fast flux sql injections.The one SQL statement is used to attack the ASP/IIS using Asprox.

Fig. 4: Phase 1 of IP Addresses for a SQL Injection Attack in Fast Flux [10]

Fig. 5: Phase 2 of IP Addresses for a SQL Injection Attack in Fast Flux [10]

*3) Compounded SQL Injection Attack: Intent of Attack* Compounded SQL Injection Attack is the mixture of the two or more attacks which attack the website and causes more serious effect then the previously discussed SQL Injections. Compounded SQL Injection has came into the place due to rapid development of prevention and detection techniques against various SQL Injections.To overcome, the malicious attackers developed a technique called compounded SQL Injection. Compounded SQL Injection is derived from the mixture of SQL Injection and other Web Application Attacks which can be detailed as follows:

1) SQL Injection + DDos Attacks :
   DDOS (Distributed Denial Of Service) is defined as the attack that is used to hang a server, exhaust the resources so that the user cannot able to access it. It can be categorized as Web Application DDOS.SQL provide us to create extremely complex queries in order to get an output in our manner. The different commands which can be used in SQL Injection in order to pursue with DDOS Attack is to encode, compress, join etc.A very little research on this topic has been done but it is a very complex and successful attack.In order to pursue with this type of attack there are the basic steps to be followed which has to be done by finding the vulnerability, preparing for the vulnerability and after that the complex code is used for the attack.The greater the number of columns and rows in the database it will be easier for the SQL DDOS attack. Hence the sample code is used to make SQL DDOS attack on the website.

```
select tab1 from (select
decode(encode(convert(compress(post)
using latin1),concat(post,post,post,post)),
sha1(concat(post,post,post,post)))
as tab1 from table_1)a;
```

Fig. 6: Sample code for the SQL Injection DDoS Attack [11]

If we find using Union SQL Injection that the website is vulnerable to the SQL Injection but we got to know that only 3rd column is vulnerable so we will try to inject the payload into the website which can be achieved as follow:

```
http://exploitable-web.com/link.php?id=1'
union select 1,2,tab1,4 from
(select decode(encode(convert
(compress(post)
using latin1),
des_encrypt
(concat(post,post,post,post),8)),
des_encrypt(sha1(concat(post,post,
post,post)),9))
as tab1 from table_1)a--
```

Fig. 7: SQL Injection in order to achieve DDoS attack

We can use the sleep command present in SQL to make connections live for long that will help to do the task. Using Sleep we can also Pool out the connection in ASP.net or many other programming languages where by default maximum 100 or 150 connections are allowed at time of 30 seconds.If, we can make our connection live using Sleep command it wont allow the server to reply other users.Hence, our DDoS attack using the SQL will be achieved.

2) SQL Injection + DNS Hijacking:
   Ex-filtration of the data using Blind SQL Attack is usually slow.So, the attackers came out with DNS attack which is much faster and less noisier than the blind Sql

Attack.DNS are more allowed than any other command to access the database and connect with the arbitrary host.The attacker main goal is to embed the SQL Query in DNS request and to capture it and makes it way onto the internet.

The term DNS Hijacking does not mean web hacking of an DNS(Domain Name Server) but it relates to the modification of the DNS entries, Exploiting the administration of the web of domain registers. When DNS Hijacking has been achieved then the second part comes into an effect which is SQL Injection attack with the DNS Look up. Conceptually, the attack would be as shown in the figure where the website used is a dummy website.

```
do_dns_lookup( (select top 1 password
from users) + '.inse6140.net' );
```

Fig. 8: DNS Look up is used for obtaining the password

The SELECT statement is used to obtain the password hash that the attacker is interested and appended a domain name which we have control to the end of it (e.g.inse61400.net) which is done with the help of DNS Hijacking.At last,we perform a DNS lookup (address-based lookup for a dummy hostname).Then we run a packet sniffer on the name server for our domain and wait for the DNS record containing our hash [12]. Below is the another example for the SQL Injection with the DNS Hijacking in real time.

```
someserver.example.com.1234
> ns.inse6140.net.53 A?
0x1234ABCD.inse6140.net
```

Fig. 9: Attack Statement using DNS Hijacking and SQL Injection

The string 0x1234ABCD here represents the password hash we hope to extract using our SELECT statement.

3) SQL Injection + XSS :

According to the manager of IBM Dewey [13] says about SQL Injection + XSS attack When you get down to the nuts and bolts of it, this is a cross-site scripting attack. SQL injection was just a vehicle to get there. In his statement he means that SQL Injection is the way for setting up an attack, the rest of the work is done by XSS (Cross Site Scripting).These attacks are known as third wave attacks as they are not typically the old way of attacking but they are the commands from hiding from Network Monitoring devices.

XSS (Cross Site Scripting) can be defined as the client side code injection attack wherein the attacker can inject malicious code into to legitimate website or an application. The script is usually inserted in the input fields of a website. After inserting the scripts are executed as it is and the role of the attacker fulfills.The figure shows the normal content of the file, the content of the file after adding the script in the input fields of the website and finally the XSS Attack with the MySQL Injection.The script will try to connect with the database of the website, hence its a difficult and complex task. As JavaScript is a client side language whereas accessing to the database is usually handled by Server side languages. If the connection is successful then the attacker will have access to the database but through client side language.It is mainly used for Extraction of the data. The implementation for inserting and modifying the data will become extremely difficult. The further modification for extraction of the data can be done using the code defined below.There are innumerable websites which are vulnerable to the XSS + SQL Injection Attack [14]. These are the complex attacks and there prevention and detection becomes really difficult as most of the websites usually uses JavaScript. The development of JavaScript is in pace such as Node.js and many more. The developers are unaware of the type of vulnerability it carries.

```
print "<html>"
print "<h1>Most recent comment</h1>"
print database.latestComment
print "</html>"
```

Fig. 10: Normal code within the website for displaying the comments [15]

```
print "<html>"
print "<h1>Most recent comment</h1>"
<iframe src=http://evil.com/xss.html>
print "</html>"
```

Fig. 11: Addition of the iframe command which is used for phishing attack(XSS Attack)

4) SQL Injection attack using Cross Domain Policies of Rich Internet application

Today majority of websites uses adobe flash and Microsoft silver light for boosting and increasing the user interactivity of the websites. Using these type of application is vulnerable to the SQL Injection and Other types of attack, if care is not taken during programming a code.They use cross domain policies to run the website.Misinterpretation and not a proper use of the Cross Domain policies give rise to the vulnerability in Rich Internet Application. Cross Domain policies is an XML file which gives permission to web client to handle data in multiple domains [16]. Cross-domain policies define the list of RIA hosting domains that are allowed to retrieve content from the content providers domain.It was first observed by the Internet Storm Center in the year of 2008. When the legitimate sites were being attacked by Asprox Injection String.It was first determining the browser being used which is either firefox or Internet Explorer. Then next set was to run the Java Script file that determines the version of the flash player used which is usually for the SQL Injection attack.

Georgios [4] in his paper describes how weak statements are written when programming for Rich Internet Appli-

```
print "<html>"
print "<h1>Most recent comment</h1>"
<script>
var connection =new
ActiveXObject("ADODB.Connection") ;
var connectionstring="Data
Source=<server>;
Initial Catalog=<catalog>;User ID=<user>;
Password=<password>;Provider=SQLOLEDB";
connection.Open(connectionstring);
var rs = new
ActiveXObject("ADODB.Recordset");
rs.Open("SELECT * FROM table",
connection);
rs.MoveFirst
while(!rs.eof)
{
document.write(rs.fields(1));
rs.movenext;
}
rs.close;
connection.close;
</script>
print "</html>"
```

Fig. 12: SQL Injection attack using the XSS

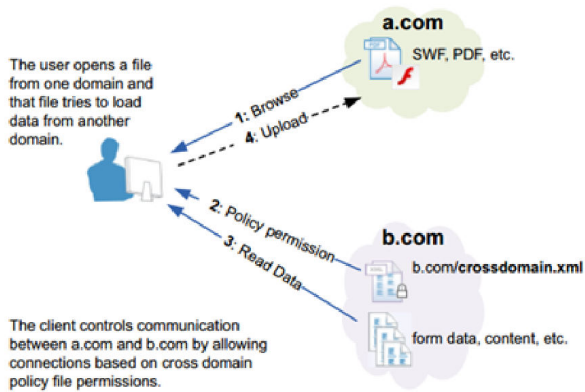

Fig. 13: Explanation of Cross Domain Policy

```
<?xml version="1.0"?>
<!DOCTYPE cross-domain-policy SYSTEM
"http://www.concordia.ca/files/docs/xml
/dtds/cross-domain-policy.dtd">
<cross-domain-policy>
<allow-access-from domain="sub1.domain1.com"/>
<allow-access-from domain="domain3.com"/>
</cross-domain-policy>
```

Fig. 14: A Valid Code for Cross Domain Policy

```
<allow-access-from domain
="*.sub1.domainA.com"/>
<allow-access-from domain="*.domainC.com"/>
<allow-access-from domain="*"/>
```

Fig. 15: A Week Code for the Cross Domain Policy

cation that is vulnerable to the attack.The first example shows the right code for the cross domain policy.If this code is written then the website is not vulnerable to SQL Injection for the Rich Internet Applications. The other figure shows if the coding style of the programmer changes which is shown in the figure then code is vulnerable to the SQL Injection Attack as well as it is vulnerable to other types of attack.There are key points in order to prevent from SQL Injection which we will discuss in Prevention and Detection section.

5) SQL Injection + Insufficient Authentication:
This type of Compounded Attack is associated with the Insufficient Authentication where the user or the site administrator is a novice.The security parameters has not been initialized where the application fails to identify the location of user, service or application.It can also refer to the website which allows the attacker to access the sensitive content without verifying the identity of the user. This advantage is taken by the attacker to induce the SQL Injection Attack. Hence, this type of attack is comparatively very easy as compared with the other types of attack. The first step is to find whether the website has insufficient authentication and if it has then the SQL Injection attack can take place.

IV. DETECTION AND PREVENTION

Detection and Prevention is a difficult task if there is proper understanding about the SQL Injection Attacks types then it is easier to prevent the attack. To prevent from modern SQL Injection Attack it is always advised to use the prepared statements [17] as it is fixed and cannot be modified by the user of a website or web application.The techniques like magic_quotes() and add_slashes() cannot protect the Web Application or Web Site from the SQL Injection Attack. Over here we will discuss various techniques for the detection and prevention of modern SQL Injection.

A. Blind SQL Injection Detection and Prevention:

There are in numerable research papers for the Blind SQL Injection where they describe different Detection and Prevention Techniques. As Blind SQL Injection are difficult to detect and prevent but researchers were aware of Blind SQL Injection from past many years. The most popular technique used is AMNESIA [18] which stands for Analysis and Monitoring for Neutralizing SQL-injection attacks.,the tool is only applicable to protect Java Based Applications and it uses run time monitoring.Komiya et al. [19] came out with the better method for preventing SQL Injection.They encouraged to use Machine Learning Algorithms in order to improve the prevention and Detection of Blind SQL Injection. They obtained the results and verified that prevention and detection were better than SQLCheck [5] and AMNESIA [18].

## B. Fast Flux SQLI Detection and Prevention:

The major attacks where the client side security fails and an emerging phenomenon which is not widely known. The fast flux SQL Injection attacks has been faced at Indiana University, US and even the security of the FBI is concerned about this type of attack.The best way to protect from fast flux attack is to make the servers safe [20]. The fast flux can be protected by using the technique by which URL's point to the Javascript delivery hits can be blacklisted if they are identified in a quick fashion. Alper et al. [21] discusses in their paper regarding the Fast Flux Monitor (FFM) that can detect as well as can classify a Fast Flux Service Networks in the order of minutes for using both active and passive DNS monitoring, which complements long term surveillance of FFSNs. After the Fast Flux Networks has been classified we can use our SQL Injection Techniques in order to stop SQL Injection which can be stopped by Monitoring.It can be possible by suggestive measures of the secure coding techniques should be taken place. As attackers are becoming smarter and finding the ways to crack into the system even the researchers came out with the new techniques to countermeasure the un-detection by the Fast Flux Monitor.Holz et al. [22] came out with a research to detect the fast flux networks and SQL Injection attacks by using the Expert Systems.The further improvement was done by Stalsman and Irwin [23] by developing more suitable method for the detection of the Fast Flux Networks and SQL Injection Attacks. They added that Machine learning methods that can be used for detection. Among many machine learning techniques they have put emphasis on C5.0 Classifier and Naive Bayesian Classifier for the detection of Fast Flux and SQL Injection Attack. Prevention of the Fast Flux is really complicated task and many researches are finding the right techniques to counter the Fast Flux SQL Injection Attacks.

## C. SQLI XSS Detection and Prevention:

Adam et al. [24] discuss in his paper about the Automation creation of the SQL Injection and XSS in order to bypass and enter into the database in order to find the vulnerability. They discuss about the Ardilla Tool which is primarily an attacking tool in which the user chooses to attack (SQLI, first order XSS or second order XSS).The tool is used for the detection of the SQLI + XSS attack.It has two modes to check the validity of the attack i.e. strict mode and lenient mode whereas SQLI has only one mode. Ardilla Tool uses Taint Based approaches and static analysis techniques, in this if the preconditions are not met ,it will suggest filters and other sanitization methods in order to fulfill the precondition which is the requirement for the detection of a vulnerability.The other tools are not as efficient as ARDILLA. Therefore, In order to protect our system firstly, XSS has to be detected and prevented. Secondly, the SQLI detection and prevention methods has to be applied in order to achieve the task.

By using the Cross Scripting Attacks the attacker can attack many different parts of the Web Application.The common being the stealing of cookies which can further lead to vulnerability, loss of critical information and SQL Injection. Stealing of cookies can be prevented by using Dynamic Cookies Rewriting techniques which has been discussed by Rattipong et al. [25].In his paper, he discuss the creation of the random data and changing of the name when storing in the cookies table. As discussed in our above section about the XSS Attack with the SQL Injection which is mostly done with the help of Java Script.In order to prevent these type of attacks Zhang et al. [26] came out with the Execution flow mechanism in order to protect from Java Script based XSS attack which serves the purpose for protecting against SQL Injection in XSS. In this prevention technique they have used the finite state automata to analyze client side java script and it prevents any malicious script to enter or retrieve the data from the database.As it uses the machine learning algorithm which improves with its experience and highly depend on the data sets but it does not guarantee full protection and it has significant performance overhead.According to vogt et al. [27] promises that Dynamic Data Tainting is a technique which is used for the detection and prevention of the XSS Attack and then SQL Injection is automatically protected but Nikiforakis et al.[28] has counter reaction as they say there are many hidden channels which remain undetected and hence cannot be prevented from attack by using the Dynamic Data Tainting.

The other tool very popular tool used to mitigate the XSS attack is the Noxes tool [29].The developers to these tool were inspired by the Windows Firewall. It has certainly helped in protection against XSS attacks + SQL Injection attacks but it fails to prevent the attack completely as discussed by Nikiforakis et al.[28] as they consider that attacker can use HTML Tags instead of Script Tags for an Attack.It takes care about HTTP request and prevents the modification done on the HTTP header and has the functionality to set cookies.

## D. SQLI DNS Detection and Prevention

In SQL + DNS Prevention and Detection, the rules of dividing will apply. In this approach, DNS Hijacking is detected and after wards SQL Injection prevention and detection takes place. DNS Hijacking can be prevented by not downloading the free utilities from the websites as they mostly contain the vulnerabilities. Diter gollman [30] describe DNS rebinding which tries to capture the router settings of the client or user.In order to prevent the DNS Hijacking the Nikiforakis et al. [28] came out with the session shield that is light weight client side protection mechanism. Stampar [17] in his paper discusses about the usage of SQLMap in protection of the SQLI + DNS Attack.The SQLMap has the feature of the DNS Ex filtration and there are many command lines specially designed for DNS prevention and detection. It is compatible with most of the SQL Database versions.

## E. SQLI Cross Domain Policies Detection and Prevention:

These types of attacks according to the Kontaxis et al. [4] can be protected by using proper policy implementation and reducing the usage of Any Subdomain Weakness and domain Weakness.The method developed by Steven and his colleagues is FlashOver [31] which is used to detect and prevent the XSS attacks in the Rich Internet Applications. As we know that if it is XSS Vulnerable then it would be SQLI vulnerable as SQL Injections can take place with the help of XSS. In their method, they have used the static as well as dynamic code analysis in order to achieve the protection of the SQLI Cross Domain Policies.In this method using the Static Analysis they retrieve the Potentially exploitable variables (PEV) which is later used as an attack vector in FLASHOVER. The method DEMACRO [32] was proposed by Sebastian and his colleagues at SAP Research Center, Germany. Their system for the prevention does not need any training or machine learning methodology.DEMACRO detects the malicious cross domain requests and tries to de-authenticate them.They did extensive research and came out with the prevention policy for the Cross Domains.

### F. SQLI DDoS Detection and Prevention:

DDoS Attacks are well understood by the security professionals but even though its well discussed topic, there are some loopholes which attacker uses to attack the system. The DDoS and SQL Injection Detection was well discussed by Lee and his colleagues which gave the idea of detecting the DDoS Attack using the cluster analysis [33]. The cluster analysis methodology helps to detect DDoS Attacks and can easily identify the type of attack on the system. Yu shui [34] came forth with the discussion of the survey of different detection techniques of the DDoS Attacks. After the detection phase mitigation comes which is comparatively much easier in the case of the SQLI DDoS attacks.Hence, the research for the DDoS SQLI is widespread . The thing needed at the present time is to utilize the techniques in a proper manner so that we can secure our web servers , web applications and websites from these type of attacks.

### G. SQLI Insufficient Authentication:

In order to protect from the SQLI without getting into the trouble of Authentication. The administrator can use the technique of Crypto-graphical Hash functions as used by Singh et al.[35] for protection from SQLI + Insufficient Authentication. In this method two extra attributes are added which are hash functions for the username and password field. The hash functions are automatically generated using Hash Algorithms. Now, when the client enters the username and password then hash function is generated and is transferred to the server side for verification. Everything which takes place over here is in encrypted form. If the username and password is same as stored in the database which is matched with its hash functions. Hence, there is negligible chance for intrusion into the database.

## V. EVALUATION

In this section we will evaluate the techniques presented in the earlier section.We have made two tables depicting in this section. In table 1. we have shown which technique is used to detect and prevent from the modern SQL Injection Attacks. In table 2. discuss about the various detection and prevention tools used.

In this section we will use asterisk (*) for showing that it is used for both Detection and Prevention in respect with the modern SQL Injections on the table.The circle (o) is used for showing that it is only used for the Detection mechanism.The plus (+) is used for depicting only prevention corresponding to the modern SQL Injection.The symbol times (x) is used to depict that techniques or tools does not correspond for the modern SQL Injection (In terms of Prevention and Detection).The symbol (p) depicts the incompleteness which means after applying the specific method the other method has to be applied in order to achieve complete detection and prevention.

### A. Evaluation on the basis of Detection and Prevention Techniques against Attack:

In the table we have taken different techniques which can be used in order to detect and prevent the attacks. These techniques will help the researchers and security professionals to take proper action or use the specified techniques to solve the crises arisen inside the organization due to an attack. The technques described here can be used to develop a system with the optional functionality in order to protect the system from any kind of these modern attacks that corresponds to the Compounded SQL Injection and Fast Flux SQL Injection attacks. According to the survey performed, we observed that the Static Code Analysis and Machine Learning are the best among the others but other techniques has various other advantages.

### B. Evaluation on the basis of Detection and Prevention Tools against Attack:

In this table we have discussed different tools for the detection and prevention from the modern attacks. These tools are ready made and some are open source which can be downloaded from the internet. Most of these tools were developed for the research purposes but due to its significant advantages they are being used in the commercial sectors. The tools are discussed giving the broad overview that which of these tools can be used for the particular type of attacks. According to our survey performed we observe that the **Noxeus** and **SQLMap** are latest and have better prevention and detection mechanism.

## VI. CHALLENGES
1) The challenge we have faced writing this survey paper is of very little scientific research being done in the field of Fast Flux SQL Injection and Compounded SQL Injection. As, it was very hard to determine proper tools for prevention and detection.
2) These topics are difficult to understand and tools should be quite sophisticated in order to find the deviation from the normal SQL statements.

| S.No. | Technique | Blind SQL Injection | FF_SQLI | SQLI_XSS | SQLI_DNS | SQLI_CDP | SQLI_DDoS | SQLI_In_Authen |
|---|---|---|---|---|---|---|---|---|
| 1. | Crypto Graphical Hash Functions | p | x | x | x | x | x | + |
| 2. | Dynamic Cookies Rewriting | x | x | + | p | x | x | p |
| 3. | Execution Flow Mechanism | x | x | * | x | x | x | x |
| 4. | Static Code Analysis | o | x | o | x | * | x | x |
| 5. | Dynamic Data Tainting | x | x | * | x | x | x | x |
| 6. | Run Time Monitoring | p | + | * | x | x | x | x |
| 7. | Machine Learning | x | o | * | x | x | o | x |

TABLE I: The different techniques for Detection and Prevention of an Attack

| S.No. | Tools | Blind SQL Injection | FF_SQLI | SQLI_XSS | SQLI_DNS | SQLI_CDP | SQLI_DDoS | SQLI_In_Authen |
|---|---|---|---|---|---|---|---|---|
| 1. | Ardilla Tool | x | x | o | x | o | x | x |
| 2. | Noxes | x | x | p | x | + | x | x |
| 3. | Session Shield | x | x | * | p | x | x | p |
| 4. | AMNESIA | * | x | p | x | x | x | x |
| 5. | SQLMap | o | x | o | p | x | o | o |
| 6. | Fast Flux Monitor | x | o | x | o | x | x | x |

TABLE II: The Evaluation of different tools for Detection and Prevention of an Attack

## VII. CONCLUSION

In our paper, we have tried to discuss the modern SQL Injection attack which are less known to the general world as well as many researchers. They are very typical attack which are done on the web applications and web sites. They take considerable amount of time to understand as they are quite complex when compared with the classical SQL Injection Attacks. We have discussed the prevention and detection techniques of these attacks which we could able to find and apply in order to prevent these attacks. The prevention and detection techniques discussed are limited due to very less research done on these types attacks.These attacks can overcome the previous detection and prevention techniques. Hence, some times proper coding of Web Application holds very little value as it can overcome easily. The developer should have the good knowledge of these type of attacks can destroy the web application and whose implication can effect the businesses of an organization.

Lastly, we have came up with the Evaluation of different detection and prevention techniques in which we compared it and came out with the general characteristics of the tools used.The future research or evaluation can be done to make more simplified and better protection and detection techniques while keeping in mind about the comparison of performance of different techniques related to Blink SQL Injection, Fast Flux SQL Injection and Compounded SQL Injection.

## APPENDIX

The technical parameters in the paper.

1) FF_SQLI : Fast Flux SQL Injection
2) SQLI_XSS : SQL Injection Attack + XSS
3) SQLI_DNS : SQL Injection + Domain Name Server
4) SQLI_CDP : SQL Injection + Cross Domain Policies
5) SQLI_DDoS : SQL Injection + DDoS Attacks
6) SQLI_In_Authen : SQL Injection + Insufficient Authentication

Run Time Monitoring : It stands for observing the program during the execution of the program. The other term used for Run time Monitoring is Run time Verification.

Static Analysis:It is a methodology in which the vulnerabilities and errors are checked without running a program.

SQLMap: it is a penetration tool used for the detection of the major SQL Injection Attacks.